\documentclass[aps,floatfix,showpacs,superscriptaddress,preprint]{revtex4}

\usepackage{graphicx,epsfig,epstopdf}
\usepackage{amssymb,amsmath,amsxtra,amsfonts}
\usepackage{bm}
\usepackage{float}

\usepackage{longtable}
\usepackage{multirow}
\usepackage{booktabs}
\usepackage{array}
\usepackage{wrapfig}
\usepackage{color}

\usepackage{titlesec}
\titleformat{\section}{\large\bfseries}{\thesection}{1em}{}

\usepackage{color}
\usepackage{slashed}
\usepackage{bbm}

\newcommand{\bea}{\begin{eqnarray}}
\newcommand{\ena}{\end{eqnarray}}
\newcommand{\be}{\begin{equation}}
\newcommand{\en}{\end{equation}}
\newcommand{\nn}{\nonumber\\}
\newcommand{\ed}{\end{document}} 
\newcommand{\Tr}{\mbox{\rm{tr}}}

\newcommand{\Jpsi}{\ensuremath{J\!/\!\psi}}

\begin{document}

\title{ $Y(4260)$ as four-quark state}

\author{S. Dubni\v{c}ka}
\email{stanislav.dubnicka@savba.sk}
\affiliation{Institute of Physics, Slovak Academy of Sciences, 
Bratislava, Slovak Republic}

\author{A. Z. Dubni\v{c}kov\'a}
\email{anna.dubnickova@fmpha.uniba.sk}
\affiliation{Dept. of Theoretical Physics, Comenius University, 
Bratislava, Slovak Republic}

\author{A. Issadykov}
\email{issadykov.a@gmail.com}
\affiliation{Bogoliubov Laboratory of Theoretical Physics, 
Joint Institute for Nuclear Research, 141980 Dubna, Russia}

\author{M. A. Ivanov}
\email{ivanovm@theor.jinr.ru}
\affiliation{Bogoliubov Laboratory of Theoretical Physics, 
Joint Institute for Nuclear Research, 141980 Dubna, Russia}

\author{A. Liptaj}
\email{andrej.liptaj@savba.sk}
\affiliation{Institute of Physics, Slovak Academy of Sciences, 
Bratislava, Slovak Republic}

\begin{abstract}
We treat the $Y(4260)$ resonance as a four-quark state in the framework of the covariant confining quark model.
We study two choices of the interpolating current, 
either the molecular-type current which effectively corresponds to
the product of $D$ and $\bar D_1$ quark currents or tetraquark one.
In both cases we  calculate the widths of decays
$Y(4260)\to Z_c(3900)+\pi$ and $Y(4260)\to D^{(\ast)}+\bar D^{(\ast)}$.
It is found that in both approches the mode $Y\to Z^+_c + \pi^-$
is enhanced compared with the open charm modes. However the absolute
value of the $Y\to Z^+_c + \pi^-$ decay width obtained in
molecular picture is arguably too large. On the other hand the value obtained in tetraquark picture is reasonable.
  
\end{abstract}
\maketitle

\section{Introduction}
\label{sec:intro}


In 2005 BABAR Collaboration observed a broad resonance around $4.26$~GeV
in analyzing the mass spectrum  of $\pi^+\pi^-\Jpsi$
in initial-state-radiation (ISR) production
$e^+e^-\to\gamma_{\rm ISR} \pi^+\pi^-\Jpsi$ \cite{Aubert:2005rm}.
Since this resonance was found in the $e^+e^-$−annihilation through ISR,
its spin-parity is $J^{PC}=1^{--}$. However, its mass does not
fit any mass of charmonium  states in the same mass
region, such as the $\psi(4040)$, $\psi(4160)$, and $\psi(4415)$.
Moreover, the $Y(4260)$ has strong coupling to the $\pi^+\pi^-\Jpsi$
final state, but no evidence was found for coupling to any
open charm decay modes as $D^{(\ast)}\bar D^{(\ast)}, D_s^{(\ast)}\bar D_s^{(\ast)} $
where $D^{(\ast)}=D$ or $D^\ast$
\cite{CroninHennessy:2008yi,Abe:2006fj,Aubert:2006mi,
  Aubert:2009aq,delAmoSanchez:2010aa}.
These properties perhaps indicate that the $Y(4260)$ state is not
a conventional state of charmonium~\cite{Brambilla:2010cs}.


In addition to the $Y(4260)$, the the BESIII Collaboration
reported on the observation of another exotic state named as  $Z_c(3900)$
in the reaction $e^+e^-\to\pi^+\pi^-\Jpsi$ \cite{Ablikim:2013mio}.
It carries an electric charge and couples to charmonium.
A fit to the $\pi^\pm\Jpsi$ invariant mass spectrum results in a mass of
$ M_{Z_c} = 3899.0 \pm 3.6({\rm stat}) \pm 4.9({\rm syst})$~MeV 
and a width of $\Gamma_{Z_c}=46 \pm 10({\rm stat}) \pm 20({\rm syst})$~MeV.
This state was confirmed by Belle~\cite{Liu:2013dau}
and CLEO~\cite{Xiao:2013iha} Collaborations. Then the BESIII Collaboration
observed  a distinct charged structure in the $ (D\bar D^\ast)^\mp $ invariant
mass distribution of the process
$e^+e^-\to\pi^\pm (D\bar D^\ast)^\mp$~\cite{Ablikim:2013xfr}. 
Assuming this structure and  the $Z_c(3900)\to \pi\Jpsi$ signal are 
from the same source, the ratio of partial widths is  
$ \Gamma(Z_c\to D\bar D^\ast)/\Gamma(Z_c\to \pi\Jpsi)= 6.2 \pm 2.7\,. $
That means that the $Z_c(3900)$ state has a much stronger coupling to $DD^\ast$
than to $\pi\Jpsi$ \cite{Liu:2015pma}.


Now we go back to the $Y(4260)$ and shortly review some theoretical
efforts to understand the underlying structure of this state.
We refer to Ref.~\cite{Brambilla:2010cs} for more complete review
of this subject. Probably, one of the first attempts to analyze the
possible interpretations of the $Y(4260)$ was undertaken
in Ref.~\cite{Zhu:2005hp}. The conclusion has been done that only 
the hybrid charmonium picture is not in conflict with available
experimental data from BABAR measurement. The interpretation
of the  $Y(4260)$  as a charmonium hybrid has been also explored in
Refs.~\cite{Kou:2005gt,Close:2005iz}.

The three-body $\Jpsi\pi\pi$ and $\Jpsi K\bar K$ systems have been treated
as coupled channels in Ref.~\cite{MartinezTorres:2009xb}. It was found
by solving the Faddeev equations that the resonance $Y(4260)$ can be
generated  due to the interaction between these three mesons.
The $Y(4260)$ has been identified as the low member of the pair
$\psi(4S)-\psi(3D)$ charmonium by using simple quark
model~\cite{LlanesEstrada:2005hz}.

In the paper~\cite{Liu:2005ay} it was suggested that the $Y(4260)$
is a $\chi_{c1}-\rho^0$ molecule. In that picture one can show that
the width  of decay $Y(4260)\to\pi^+\pi^-\Jpsi$ is larger than
$Y(4260)\to D\bar D$  which has not been observed.

It was proposed in Ref.~\cite{Maiani:2005pe} to interpret the $Y(4260)$
as the first orbital excitation of a diquark-antidiquark state
$([cs][\bar c \bar s])$. In this case the $Y(4260)$  should decay predominantly
to $D_s\bar D_s$. 

Masses of heavy tetraquarks have been calculated
in the relativistic quark model \cite{Ebert:2005nc}.
It was found the P-wave state of the tetraquark
combination $(([cq]_{S=0}[\bar c \bar q]_{S=0})$ has a mass of 4244~MeV
which is close to the $Y(4260)$ mass. At the same time the mass of
charm-strange diquark-antidiquark was found to be more than 200~MeV
heavier than the $Y(4260)$ mass. It was concluded that
a more natural tetraquark interpretation of the  $Y(4260)$
is charm-nonstrange diquark-antidiquark state. Then the dominant decay
mode of the $Y(4260)$ would be in $D\bar D$ pairs.

However, as mentioned above, no evidence was found for the decays
$Y(4260)\to D^{(\ast)}\bar D^{(\ast)}, D_s^{(\ast)}\bar D_s^{(\ast)} $
\cite{CroninHennessy:2008yi,Abe:2006fj,Aubert:2006mi,
  Aubert:2009aq,delAmoSanchez:2010aa}.
In the Ref.~\cite{Wang:2013cya} it was assumed that the  $Y(4260)$
is $D\bar D_1$ molecular state where $D=D(1870)$ is the psedoscalar meson
with the quantum numbers $I(J^P)=\frac12 (0^-)$ and $D_1=D_1(2420)$ is
the narrow axial meson  $I(J^P)=\frac12 (1^+)$,  $\Gamma = 27\pm 3$~MeV.
With this ansatz, the observation of $Z_c(3900)$ in the $\pi^+\pi^-\Jpsi$ 
invariant mass distribution has as obvious explanation
as well the absence of the $Y(4260)$ in the decays with open charm.

However, in Ref.~\cite{Li:2013yka} it was argued that the production of
an S-wave $DD_1$ pair in $\ell^+\ell^-$−annihilation is forbidden
by the heavy quark spin symmetry. This argument is certainly not in the favor
of considering the $Y(4260)$ as  S-wave $DD_1$ state. Despite of this,
there are many studies of the $Y(4260)$ as $DD_1$ molecular state.
We briefly mention some of them.
By assuming that the $Y(4260)$ is a $DD_1$ molecular state, some hidden-charm
and charmed pair decay channels of the $Y(4260)$ via intermediate
$DD_1$ meson loops within an effective Lagrangian approach have been
investigated in Ref.~\cite{Li:2013yla}. By treating the $Y(4260)$ as
a $DD_1$ weakly bound state and also the $Z_c(3900)$ as a $DD^\ast$ molecule
\cite{Dong:2013kta}, the two-body decay $Y(4260)\to Z_c(3900) + \pi$ has been
studied. Moreover the decay mode $Y(4260) \to \Jpsi +\pi^+\pi^- $ was also
computed.

The approach we propose is based on the covariant confining quark model (CCQM) 
~\cite{Efimov:1988yd, Efimov:1993zg, Branz:2009cd} which represents an effective quantum field treatment of hadronic effects.
The model is derived from Lorentz invariant non-local Lagrangian in which a hadron is coupled to its constituent quarks. Hadrons are characterized by size parameters $\Lambda_H$ from which the strength of the quark-hadron coupling can derived. It is done by using so-called compositeness condition ~\cite{Salam:1962ap, Weinberg:1962hj}, this condition requires the wavefunction renormalization constant of the hadron to be zero $Z_H=0$. Besides reducing the number of free parameters (i.e. couplings), it also guarantuees a correct description of bound states as dressed (with no overlap with bare states) and solves the double counting problem. The vertices are  described by a Gaussian-type 
vertex functions which are supposed to effectively include contributions from gluons (which are not present). Thanks to the built-in confinement, based on a cutoff in the integration space of Schwinger parameters (stemming from representation of quark proparators), the model can be used for description of arbitrary heavy hadrons. The model should be understood as a practical tool for computing hadronic form factors from assumed quark currents, which is, in this text, applied to  $Y(4260)$ and $Z_c(3900)$ states.

In our earlier papers devoted to description of the multi-quark states
Refs.\cite{Dubnicka:2010kz,Dubnicka:2011mm}, first, we have explored
the consequences of treating the $X(3872)$ meson as a tetraquark,
i.e. diquark-antidiquark bound state. We have calculated the decay widths of
the observed channels and concluded that for reasonable values of the size
parameter of the $X(3872)$ one finds consistency with the available
experimental data. Then we have critically checked in Ref.~\cite{Goerke:2016hxf}
the tetraquark picture for the $Z_c(3900)$ state by analyzing its strong
decays. We found that $Z_c(3900)$ has a much more stronger coupling to
$DD^\ast$ than to $\Jpsi\pi$ which is in discord with experiment.
As an alternative we have employed a molecular-type four-quark current to
describe the decays of the $Z_c (3900)$ state. We found that a molecular-type
current gives the values of the above decays in accordance with
the experimental observation. By using molecular-type four-quark currents for
the recently observed resonances $Z_b(10610)$ and $Z_b(10650)$, we have
calculated in Ref.~\cite{Goerke:2017svb} their two-body decay rates into
a bottomonium state plus a light meson as well as into B-meson pairs.
A brief sketch of our findings may be found in Ref.~\cite{Ivanov:2018ayq}.

In the present paper we treat the $Y(4260)$ resonance as
a four-quark state. We study two choices of the interpolating currents
either the molecular-type current which effectively corresponds to the product
of $D$ and $\bar D_1$ quark currents or tetraquark one.
In both cases we calculate the widths of decays
$Y(4260)\to Z_c(3900)+\pi$ and $Y(4260)\to D^{(\ast)}+\bar D^{(\ast)}$.

The paper is organized as follows: Two subsequent sections~\ref{sec:Ymol} and~\ref{sec:Y4q}  are dedicated to the general formalism for describing $Y(4260)$ as four quark molecular state and tertaquark state respectively, full expressions of studied quark currents and related amplitudes are provided. In the next, last section the decay width formulas are written down and used to reach our numerical results which are presented together with our conclusion.

\section{Y(4260) as four-quark state with molecular-type current \label{sec:Ymol}}

We start with an assumption that both the $Y(4260)$ and $Z^+_c(3900)$
resonances are four-quark states with the molecular-type currents given
in Table~\ref{tab:mol}.

\begin{table}[H]
\caption{Quantum numbers and molecular-type currents.}
\centering
\vspace*{3mm}
\def\arraystretch{1.5}
\begin{tabular}{c|c|c|c|c}
\hline
Title  & $I^G(J^{PC})$  &  Interpolating current  & Mass (MeV) & Width (MeV) \\
\hline
$Y(4260)$  & $ 0^-(1^{--})$ &
$\tfrac{1}{\sqrt{2}}\Big\{
(\bar q\,\gamma_5 \, c)(\bar c\, \gamma^\mu\gamma_5 q)
- (\gamma_5\leftrightarrow \gamma^\mu\gamma_5)\Big\}$ & 4230$\pm$8 & 55$\pm$19
\\
$Z^+_c(3900) $ & $1^+(1^{+-})$ &
$ \tfrac{i}{\sqrt{2}}\Big\{ (\bar d\,\gamma_5 \, c)(\bar c\, \gamma^\mu u)
   +  (\gamma_5\leftrightarrow \gamma^\mu)\Big\}$ &
3887.2$\pm$2.3 & 28.2$\pm$2.6
\\
\hline
\end{tabular} 
\label{tab:mol}
  \end{table}
Their nonlocal generalizations are given by
\bea
J^\mu_{Y^{\rm mol}}(x) &=& \int\! dx_1\ldots \int\! dx_4 
\delta\left(x-\sum\limits_{i=1}^4 w_i x_i\right) 
\Phi_{\,Y}\Big(\sum\limits_{i<j} (x_i-x_j)^2 \Big)
J^\mu_{Y^{\rm mol};4q}(x_1,\ldots,x_4),
\label{eq:Y-mol}\\
J^\mu_{Y^{\rm mol};4q}&=&  \tfrac{1}{\sqrt{2}} 
\Big\{
 (\bar q(x_3) \gamma_5   c(x_1))\cdot (\bar c(x_2) \gamma^\mu\gamma_5 q(x_4) )
-(\gamma_5\leftrightarrow \gamma^\mu\gamma_5)\Big\} \quad (q=u,d).
\nn[1.2ex]
J^\mu_{Z^{\rm mol}_c}(x) &=& \int\! dx_1\ldots \int\! dx_4 
\delta\left(x-\sum\limits_{i=1}^4 w_i x_i\right) 
\Phi_{\,Z}\Big(\sum\limits_{i<j} (x_i-x_j)^2 \Big)
J^\mu_{Z^{\rm mol};4q}(x_1,\ldots,x_4),
\label{eq:Z-cur}\\
J^\mu_{Z^{\rm mol};4q}&=&  \tfrac{i}{\sqrt{2}} 
\Big\{
 (\bar d(x_3) \gamma_5   c(x_1))\cdot (\bar c(x_2) \gamma^\mu u(x_4) )
+ (\gamma_5\leftrightarrow \gamma^\mu)\Big\}.
\nonumber
\ena 
The reduced quark masses are specified as
\be
w_1=w_2 = \frac{m_c}{2(m_c+m_q)}, \qquad w_3=w_4 = \frac{m_q}{2(m_c+m_q)},
\en
where we assume no isospin-violation in the $u-d$ sector, i.e.
$m_u=m_d$. The Fourier-transform of the vertex function $\Phi$ may be written
as
\be
\Phi\Big(\sum\limits_{i<j} (x_i-x_j)^2 \Big) =
\prod\limits_{i=1}^3\int\!\!\frac{d^4 q_i}{(2\pi)^4}\,
e^{-iq_1(x_1-x_4)-iq_2(x_2-x_4)-iq_3(x_3-x_4)}
\widetilde\Phi\left(-\,\frac12\sum\limits_{i\le j}q_iq_j\right). 
\en

\begin{figure}[ht]
\begin{center}
\epsfig{figure=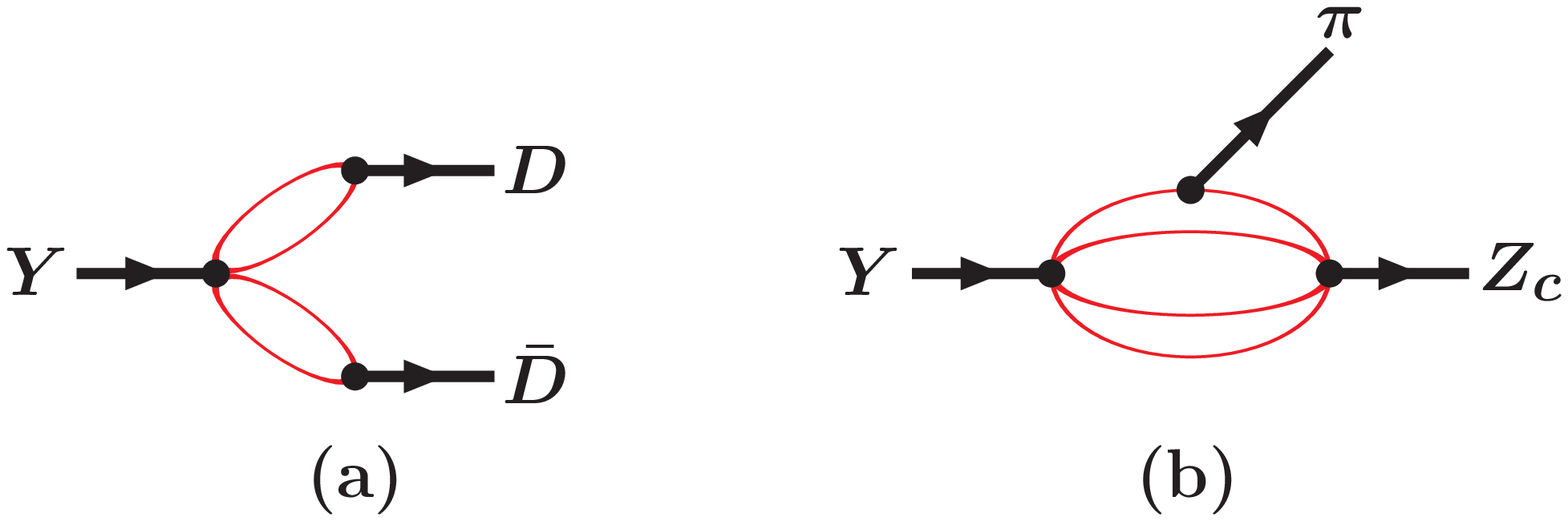,scale=.65}
\vspace*{-.25cm}
\caption{Two modes of the $Y(4260)$ decay.}
\label{fig:Y-decay}
\end{center}
\end{figure}

We consider two kinds of the strong $Y$-decays:
$Y\to D+\bar D$ where we imply the open-charm combinations as
$D\bar D$, $D\bar D^\ast$, $D^\ast\bar D$, $D^\ast\bar D^\ast$,
and $Y\to Z_c+\pi$. The Feynman diagrams describing these decays
are shown in Fig.~\ref{fig:Y-decay}.
The matrix elements of the decays $Y_u\to D_1 + \bar D_2$ read as
\bea
&&
M\left( Y_u(p,\epsilon^\mu_p)\to D_1^0(p_1) + \bar D_2^0(p_2)\right)
= \frac{9}{\sqrt{2}}\,g_{Y}g_{D_1}g_{D_2}
\nn
&\times&
\int\!\!\frac{d^4k_1}{(2\pi)^4i}\,\int\!\!\frac{d^4k_2}{(2\pi)^4i}\,
\widetilde\Phi_{Y}\left(-\,\Omega_q^{\,2}\right)
   \widetilde\Phi_{D_1}\left(-\,\ell_1^2\right)
\widetilde\Phi_{D_2}\left(-\,\ell_2^2\right)
\nn
&\times& \Big\{
\Tr\left[ \gamma_5 S_c(k_1)\Gamma_2 S_u(k_3)  \right]  
    \cdot\Tr\left[ \gamma^\mu\gamma_5 S_u(k_2)\Gamma_1 S_c(k_4)  \right]
-(\gamma_5\leftrightarrow \gamma^\mu\gamma_5)\Big\}.
\ena
Here, $\Gamma_1\otimes\Gamma_2 = \gamma_5\otimes\gamma_5$ for $D\bar D$ pair,
$\epsilon^\ast_{\nu_1}\gamma^{\nu_1}\otimes\gamma_5$ for $D^\ast\bar D$ pair,
and $\epsilon^\ast_{\nu_1}\gamma^{\nu_1}\otimes\epsilon^\ast_{\nu_2}\gamma^{\nu_2}$
for $D^\ast\bar D^\ast$ pair. The momenta are defined as
\bea
\Omega^2_q &=& \frac12 \sum\limits_{i\le j} q_iq_j,
\quad q_1=-k_1-w_1^Y p, \quad q_2=k_4-w_2^Y p, \quad q_3=k_3-w_3^Y p,
\nn
\ell_1&=&k_2+w^D_u p_1, \quad \ell_2 =-k_1-w^D_c p_2,
\quad k_3=k_1+p_2, \quad k_4=k_2+p_1.
\ena
The calculation of the matrix element of the decay $Y\to Z_c+\pi$
is more involved because it is described by three-loop diagram as shown
in Fig~\ref{fig:Y-decay}~b. One has
\bea
&&
M\left( Y_u(p,\epsilon^\mu)
\to Z_c^+(p_1,\epsilon^\nu) + \pi^-\right)
=\frac{9}{2}\,g_{Y}g_{Z_c}g_{\pi}
\nn
&\times&
\prod\limits_{j=1}^3\Big[\int\!\!\frac{d^4k_j}{(2\pi)^4i}\Big]\,
\widetilde\Phi_{Y}\left(-\,\Omega_q^{\,2}\right)
   \widetilde\Phi_{Z_c}\left(-\,\Omega_r^2\right)
\widetilde\Phi_{\pi}\left(-\,\ell^2\right)
\nn
&\times&
\epsilon_\mu(p)\epsilon^\ast_\nu(p_1)\sum\limits_{\Gamma}
\,\,\Tr\left[ \Gamma_1 S_c(k_1)\Gamma_2 S_u(k_2)  \right]  
    \cdot\Tr\left[ \Gamma_3 S_u(k_3)\Gamma_4 S_d(k_4)\Gamma_5 S_c(k_5)\right]
\,.
\ena
Here
\bea
&&
\sum\limits_{\Gamma} \left[\Gamma_1\otimes\Gamma_2\right]\cdot
                     \left[\Gamma_3\otimes\Gamma_4\otimes\Gamma_5\right]
= \left[\gamma_5\otimes\gamma_5\right]\cdot
\left[\gamma^\mu\gamma_5\otimes\gamma_5\otimes\gamma^\nu\right]
\nn
&-& \left[\gamma^\mu\gamma_5\otimes\gamma^\nu\right]\cdot
    \left[\gamma_5\otimes\gamma_5\otimes\gamma_5\right]
- \left[\gamma^\mu\gamma_5\otimes\gamma_5\right]\cdot
  \left[\gamma_5\otimes\gamma_5\otimes\gamma^\nu\right],
\ena
The momenta are defined as
\bea
\Omega^2_q &=& \frac12 \sum\limits_{i\le j} q_iq_j,
\quad q_1=-k_1-w_1^Y p, \quad q_2=k_5-w_2^Y p, \quad q_3=k_2-w_3^Y p,
\nn
\Omega^2_r &=& \frac12 \sum\limits_{i\le j} r_ir_j,
\quad r_1=-k_5+w_1^Z p_1, \quad r_2=k_1+w_2^Z p_1, \quad r_3=k_4-w_3^Z p_1,
\nn
\ell&=&k_3+w^\pi_u p_2, \quad k_4=k_3+p_2, \quad k_5=k_1-k_2+k_3+p.
\ena

\section{Y(4260) as four-quark state with tetraquark current \label{sec:Y4q}}

Now we treat the $Y(4260)$ as four-quark state with the tetraquark
current:
\be
J^\mu_{Y^{\rm tet}} = \tfrac{1}{\sqrt{2}}\,\epsilon_{abc}\epsilon_{dec}
\Big\{ (q_a\,C\gamma_5 \, c_b)(\bar q_d\, \gamma^\mu\gamma_5C \bar c_e)
-  (q_a\,C\gamma^\mu\gamma_5 \, c_b)(\bar q_d\,\gamma_5C \bar c_e). 
\en
where the charge conjugate matrix is chosen in the form $C=\gamma^0\gamma^2$
so that $C^T = - C$, $C^\dagger = C$ and $C^2=C$.
Its nonlocal generalization is given by
\bea
J^\mu_{Y^{\rm tet}}(x) &=& \int\! dx_1\ldots \int\! dx_4 
\delta\left(x-\sum\limits_{i=1}^4 w^Y_i x_i\right) 
\Phi_{\,Y}\Big(\sum\limits_{i<j} (x_i-x_j)^2 \Big)
J^\mu_{Y^{\rm tet};4q}(x_1,\ldots,x_4),
\nn
J^\mu_{Y^{\rm tet};4q} &=&  \tfrac{1}{\sqrt{2}}\, \epsilon_{abc}\epsilon_{dec}
\Big\{
 (q_a(x_4) C\gamma_5 c_b(x_1))(\bar q_d(x_3) \gamma^\mu\gamma_5C \bar c_e(x_2))
-(\gamma_5\leftrightarrow\gamma^\mu\gamma_5)\Big\}.
\ena

The matrix elements of the decays $Y_u\to D_1 + \bar D_2$ read as
\bea
&&
M\left( Y^{\rm tet}_u(p,\epsilon^\mu_p) \to D_1^0(p_1) + \bar D_2^0(p_2)\right)
=\frac{6}{\sqrt{2}}\,g_{Y}g_{D_1}g_{D_2}
\nn
&\times&
\int\!\!\frac{d^4k_1}{(2\pi)^4i}\,\int\!\!\frac{d^4k_2}{(2\pi)^4i}\,
\widetilde\Phi_{Y}\left(-\,\Omega_q^{\,2}\right)
   \widetilde\Phi_{D_1}\left(-\,\ell_1^2\right)
\widetilde\Phi_{D_2}\left(-\,\ell_2^2\right)
\nn
&\times& \Big\{
Tr\left[ \gamma_5 S_c(k_1)\Gamma^D_2 S_u(k_3)   
   \gamma^\mu\gamma_5 S_c(k_2)\Gamma^D_1 S_u(k_4)  \right]
-( \gamma_5\leftrightarrow\gamma^\mu\gamma_5 )
\Big\}\,.
\ena
The momenta are defined as
\bea
\Omega^2_q &=& \frac12 \sum\limits_{i\le j} q_iq_j,
\quad q_1=-k_1-w_1^Y p, \quad q_2=-k_2-w_2^Y p, \quad q_3=k_3-w_3^Y p,
\nn
\ell_1&=&-k_2-w^D_c p_1, \quad \ell_2 =-k_1-w^D_c p_2,
\quad k_3=k_1+p_2, \quad k_4=k_2+p_1.
\ena
The matrix element of the decay $Y\to Z_c+\pi$ is written down
\bea
&&
M\left( Y^{\rm tet}_u(p,\epsilon^\mu)
\to Z_c^+(p_1,\epsilon^\nu) + \pi^-(p_2)\right)
= 3\,g_{Y}g_{Z_c}g_{\pi}
\nn
&\times&
\prod\limits_{j=1}^3\Big[\int\!\!\frac{d^4k_j}{(2\pi)^4i}\Big]\,
\widetilde\Phi_{Y}\left(-\,\Omega_q^{\,2}\right)
   \widetilde\Phi_{Z_c}\left(-\,\Omega_r^2\right)
\widetilde\Phi_{\pi}\left(-\,\ell^2\right)
\nn
&\times&
\epsilon_\mu(p)\epsilon^\ast_\nu(p_1)\sum\limits_{\Gamma}
\,\,\Tr\left[ \Gamma^Y_1 S_c(k_1)\Gamma^Z_2 S_u(k_2)    
    \Gamma_2^Y S_c(k_3)\bar\Gamma^Z_1 S_d(k_4)\gamma_5 S_u(k_5)\right]
\ena
where $\bar\Gamma = C^{-1}\Gamma^T C$ and
$
\sum\limits_{\Gamma} =
\left[
  \gamma_5\otimes\gamma^\mu\gamma_5 -\gamma^\mu\gamma_5\otimes \gamma_5
  \right]^Y
\otimes
\left[
  \gamma_5\otimes\gamma^\nu -\gamma^\nu\otimes \gamma_5
  \right]^Z.
$
The momenta are defined as
\bea
\Omega^2_q &=& \frac12 \sum\limits_{i\le j} q_iq_j,
\quad q_1=-k_1-w_1^Y p, \quad q_2=-k_3-w_2^Y p, \quad q_3=k_2-w_3^Y p,
\nn
\Omega^2_r &=& \frac12 \sum\limits_{i\le j} r_ir_j,
\quad r_1=k_3+w_1^Z p_1, \quad r_2=k_1+w_2^Z p_1, \quad r_3=-k_4+w_3^Z p_1,
\nn
\ell&=&-k_4-w^\pi_d p_2, \quad k_4=k_1-k_2+k_3+p_1, \quad k_5=k_1-k_2+k_3+p.
\ena

\section{Numerical results and conclusion}

We remind the formulas for the two-body decay widths expressed via
Lorentz form factors.
\bea
&&
M(V(p)\to P(p_1)+P(p_2)) = \epsilon_V^\mu q_\mu G_{VPP}\,,\qquad q=p_1-p_2,
\nn
&&
\Gamma(V\to PP) = \frac{\mathbf{|p_1|}^3}{6\pi m^2}G^2_{VPP},
\nn[1.2ex]
&&
M(V(p)\to A(p_1)+P(p_2)) = \epsilon_V^\mu\epsilon^{\ast\,\nu}_A
\left(g_{\mu\nu} A + p_{1\,\mu}p_{\nu} B\right),
\nn
&&
\Gamma(V\to AP) = \frac{\mathbf{|p_1|}}{24\pi m^2}
\Big\{\Big( 3+\frac{\mathbf{|p_1|}^2}{m_1^2}\Big) A^2
+ \frac{m^2}{m_1^2}\,\mathbf{|p_1|}^4 B^2
+ \frac{m^2+m_1^2-m_2^2}{m_1^2}\,\mathbf{|p_1|}^2 AB
\Big\},
\nn[1.2ex]
&&
M(V(p)\to V(p_1)+P(p_2)) = \epsilon_V^\mu\epsilon^{\ast\,\nu_1}_V
\varepsilon_{\mu\nu_1\alpha\beta}p^\alpha p_1^\beta G_{VVP},
\nn
&&
\Gamma(V\to VP) = \frac{\mathbf{|p_1|}^3}{12\pi} G^2_{VVP},
\nn[1.2ex]
&&
M(V(p)\to V(p_1)+V(p_2)) =
\epsilon_V^\mu\epsilon^{\ast\,\nu_1}_V\epsilon^{\ast\,\nu_2}_V
\Big\{
p_{1\,\mu}p_{1\,\nu_2}p_{2\,\nu_1} A + g_{\mu\nu_1}p_{1\,\nu_2} B
\nn
&&
\phantom{M(V(p)\to V_1(p_1)+V_2(p_2)) =
\epsilon_V^\mu\epsilon^{\ast\,\nu_1}_{V_1}\epsilon^{\ast\,\nu_2}_{V_2}}
  + g_{\mu\nu_2}p_{2\,\nu_1} C + g_{\nu_1\nu_2}p_{1\,\mu} D
  \Big\},
  \nn
  &&
  \Gamma(V\to V_1V_2) =  \frac{\mathbf{|p_1|}^3}{24\pi m_1^2m_2^2}
    \Big\{
    m^2 \mathbf{|p_1|}^4 A^2 + [ \mathbf{|p_1|}^2 -3m_1^2 ] B^2
    +  [ \mathbf{|p_1|}^2 + 3m_2^2 ] C^2
    \nn
    &&
  +  [ \mathbf{|p_1|}^2 + 3\frac{m_1^2 m_2^2}{m^2} ] D^2
  +  \mathbf{|p_1|}^2 [m^2+m_1^2 -m_2^2] AB
  \nn
  &&
  +  \mathbf{|p_1|}^2 [-m^2+m_1^2 -m_2^2] AC
  +  \mathbf{|p_1|}^2 [m^2-m_1^2 -m_2^2] AD
  \nn
  &&
  +   [2 \mathbf{|p_1|}^2 - m^2 + m_1^2 + m_2^2] BC
  +   [2 \mathbf{|p_1|}^2 + m_1^2 + \frac{m_1^2}{m^2}(m_2^2 - m_1^2) ]BD
  \nn
  &&
  +   [-2 \mathbf{|p_1|}^2 - m_2^2 + \frac{m_2^2}{m^2}(m_2^2 - m_1^2) ]CD  
  \Big\}.
\ena

We calculate the decay widths and put their numerical values
in Table~\ref{tab:numerics}. We have taken the value of $Z_c$ size
parameter to be equal $\Lambda_{Z_c}=3.3$~GeV as was obtained in our
paper~\cite{Goerke:2016hxf}. We vary the value of $Y$ size parameter
in some vicinity of this average value $\Lambda_Y=3.3\pm 0.1$~GeV.
One can see that in both approches the mode $Y\to Z^+_c + \pi^-$
is enhanced compared with the open charm modes.
The two approaches differ in the decay width values $\Gamma(Y \to Z^+_c  \pi^-)$. Comparison with the total decay width of the $Y(4260)$ particle from experiment $55 \pm 19 \mathrm{MeV}$ ~\cite{Tanabashi:2018oca} disqualifies the molecular picture. As a result, one can conclude that the CCQM model calculations favor the tetraquark picture of the $Y(4260)$ state since it leads to reasonable number of the decay width into $Z^+_c  \pi^-$.

\begin{table}[H]
\caption{Decay widths in MeV.}
\centering
\vspace*{3mm}
\def\arraystretch{1.2}
\begin{tabular}{c|c|c}
\hline
Mode            &  Molecular-type current  & Tetraquark current \\
\hline
$Y\to Z^+_c + \pi^-$  & $ 146 \pm 13 $ &
                        $  5.77 \pm 0.39 $ \\
$Y\to D^0 + \bar D^0$  & $ 11 \pm 2 $ &
                        $ (0.42 \pm 0.16)\cdot 10^{-3} $ \\
$Y\to D^{\ast\,0} + \bar D^0$ & $ (0.39\pm 0.14)\cdot 10^{-2} $ &
                               $ 0.32 \pm 0.09 $ \\ 
$Y\to D^{\ast\,0} + \bar D^{\ast\,0}$ & 0 & $ (0.19 \pm 0.08)\cdot 10^{-3} $ \\

\hline
\end{tabular} 
\label{tab:numerics}
  \end{table}

\begin{acknowledgments}
The work was supported by the Joint Research Project of Institute of Physics, Slovak Academy of Sciences (SAS), and Bogoliubov Laboratory of Theoretical Physics, Joint Institute for Nuclear Research (JINR),Grant No. 01-3-1135-2019/2023.
A. Z. Dubni\v{c}kov\'a., S. Dubni\v{c}ka and A. Liptaj also acknowledge the support from Slovak Grant Agency for Sciences (VEGA), Grant No.2/0153/17.
\end{acknowledgments}

\ed
\begin{thebibliography}{99}

  
\bibitem{Aubert:2005rm} 
  B.~Aubert {\it et al.} [BaBar Collaboration],
  Phys.\ Rev.\ Lett.\  {\bf 95}, 142001 (2005)
  [hep-ex/0506081].


\bibitem{CroninHennessy:2008yi} 
  D.~Cronin-Hennessy {\it et al.} [CLEO Collaboration],
  Phys.\ Rev.\ D {\bf 80}, 072001 (2009)
  [arXiv:0801.3418 [hep-ex]].

\bibitem{Abe:2006fj} 
  G.~Pakhlova, K.~Abe {\it et al.} [Belle Collaboration],
  Phys.\ Rev.\ Lett.\  {\bf 98}, 092001 (2007)
  [hep-ex/0608018].

\bibitem{Aubert:2006mi} 
  B.~Aubert {\it et al.} [BaBar Collaboration],
  Phys.\ Rev.\ D {\bf 76}, 111105 (2007)
  [hep-ex/0607083].

\bibitem{Aubert:2009aq} 
  B.~Aubert {\it et al.} [BaBar Collaboration],
  Phys.\ Rev.\ D {\bf 79}, 092001 (2009)
  [arXiv:0903.1597 [hep-ex]].

\bibitem{delAmoSanchez:2010aa} 
  P.~del Amo Sanchez {\it et al.} [BaBar Collaboration],
  Phys.\ Rev.\ D {\bf 82}, 052004 (2010)
  [arXiv:1008.0338 [hep-ex]].

\bibitem{Brambilla:2010cs} 
  N.~Brambilla {\it et al.},
  Eur.\ Phys.\ J.\ C {\bf 71}, 1534 (2011)
  [arXiv:1010.5827 [hep-ph]].
  
  
\bibitem{Ablikim:2013mio} 
  M.~Ablikim {\it et al.} [BESIII Collaboration],
  Phys.\ Rev.\ Lett.\  {\bf 110}, 252001 (2013)
  doi:10.1103/PhysRevLett.110.252001
  [arXiv:1303.5949 [hep-ex]].


\bibitem{Liu:2013dau} 
  Z.~Q.~Liu {\it et al.} [Belle Collaboration],
  Phys.\ Rev.\ Lett.\  {\bf 110}, 252002 (2013). 

\bibitem{Xiao:2013iha} 
  T.~Xiao, S.~Dobbs, A.~Tomaradze and K.~K.~Seth,
  Phys.\ Lett.\ B {\bf 727}, 366 (2013). 

\bibitem{Ablikim:2013xfr} 
  M.~Ablikim {\it et al.} [BESIII Collaboration],
  Phys.\ Rev.\ Lett.\  {\bf 112}, 022001 (2014). 

\bibitem{Liu:2015pma} 
  Z.~Liu,
  arXiv:1504.06102 [hep-ex].

  
\bibitem{Zhu:2005hp} 
  S.~L.~Zhu,
  Phys.\ Lett.\ B {\bf 625}, 212 (2005)
  [hep-ph/0507025].

\bibitem{Kou:2005gt} 
  E.~Kou and O.~Pene,
  Phys.\ Lett.\ B {\bf 631}, 164 (2005)
  [hep-ph/0507119].

\bibitem{Close:2005iz} 
  F.~E.~Close and P.~R.~Page,
  Phys.\ Lett.\ B {\bf 628}, 215 (2005)
  [hep-ph/0507199].

\bibitem{MartinezTorres:2009xb} 
  A.~Martinez Torres, K.~P.~Khemchandani, D.~Gamermann and E.~Oset,
  Phys.\ Rev.\ D {\bf 80}, 094012 (2009)
  [arXiv:0906.5333 [nucl-th]].
  
\bibitem{LlanesEstrada:2005hz} 
  F.~J.~Llanes-Estrada,
  Phys.\ Rev.\ D {\bf 72}, 031503 (2005)
  [hep-ph/0507035].

\bibitem{Liu:2005ay} 
  X.~Liu, X.~Q.~Zeng and X.~Q.~Li,
  Phys.\ Rev.\ D {\bf 72}, 054023 (2005)
  [hep-ph/0507177].
  
\bibitem{Maiani:2005pe} 
  L.~Maiani, V.~Riquer, F.~Piccinini and A.~D.~Polosa,
  Phys.\ Rev.\ D {\bf 72}, 031502 (2005)
  [hep-ph/0507062].
  
\bibitem{Ebert:2005nc} 
  D.~Ebert, R.~N.~Faustov and V.~O.~Galkin,
  Phys.\ Lett.\ B {\bf 634}, 214 (2006)
  [hep-ph/0512230].


\bibitem{Wang:2013cya} 
  Q.~Wang, C.~Hanhart and Q.~Zhao,
  Phys.\ Rev.\ Lett.\  {\bf 111}, no. 13, 132003 (2013)
  [arXiv:1303.6355 [hep-ph]].


\bibitem{Li:2013yka} 
  X.~Li and M.~B.~Voloshin,
  Phys.\ Rev.\ D {\bf 88}, no. 3, 034012 (2013)
  [arXiv:1307.1072 [hep-ph]].

  
\bibitem{Li:2013yla} 
  G.~Li and X.~H.~Liu,
  Phys.\ Rev.\ D {\bf 88}, no. 9, 094008 (2013)
  [arXiv:1307.2622 [hep-ph]].

\bibitem{Dong:2013kta} 
  Y.~Dong, A.~Faessler, T.~Gutsche and V.~E.~Lyubovitskij,
  Phys.\ Rev.\ D {\bf 89}, no. 3, 034018 (2014)
  [arXiv:1310.4373 [hep-ph]]. 







\bibitem{Efimov:1988yd} 
  G.~V.~Efimov and M.~A.~Ivanov,
  Int.\ J.\ Mod.\ Phys.\ A {\bf 4}, 2031 (1989).
  
\bibitem{Efimov:1993zg} 
  G.~V.~Efimov and M.~A.~Ivanov,
  {\it The Quark Confinement Model of Hadrons} (CRC Press, Boca Raton, 1993). 
  
\bibitem{Branz:2009cd} 
  T.~Branz, A.~Faessler, T.~Gutsche, M.~A.~Ivanov, J.~G.~K\"{o}rner, and V.~E.~Lyubovitskij,
  Phys.\ Rev.\ D {\bf 81}, 034010 (2010)
  [arXiv:0912.3710].  




\bibitem{Salam:1962ap} 
  A.~Salam,
  Nuovo Cimento  {\bf 25}, 224 (1962).

\bibitem{Weinberg:1962hj} 
  S.~Weinberg,
  Phys.\ Rev.\  {\bf 130}, 776 (1963).




\bibitem{Dubnicka:2010kz} 
  S.~Dubni\v{c}ka, A.~Z.~Dubni\v{c}kov\'a, M.~A.~Ivanov and J.~G.~K\"orner,
  Phys.\ Rev.\ D {\bf 81}, 114007 (2010)
  [arXiv:1004.1291 [hep-ph]].

\bibitem{Dubnicka:2011mm} 
  S.~Dubni\v{c}ka, A.~Z.~Dubni\v{c}kov\'a, M.~A.~Ivanov, J.~G.~K\"orner,
  P.~Santorelli and G.~G.~Saidullaeva,
  Phys.\ Rev.\ D {\bf 84}, 014006 (2011)
  [arXiv:1104.3974 [hep-ph]].
  
\bibitem{Goerke:2016hxf} 
  F.~Goerke, T.~Gutsche, M.~A.~Ivanov, J.~G.~K\"orner, V.~E.~Lyubovitskij
  and P.~Santorelli,
  Phys.\ Rev.\ D {\bf 94}, no. 9, 094017 (2016)
  [arXiv:1608.04656 [hep-ph]].

\bibitem{Goerke:2017svb} 
  F.~Goerke, T.~Gutsche, M.~A.~Ivanov, J.~G.~K\"orner and V.~E.~Lyubovitskij,
  Phys.\ Rev.\ D {\bf 96}, no. 5, 054028 (2017)
  [arXiv:1707.00539 [hep-ph]].

\bibitem{Ivanov:2018ayq} 
  M.~Ivanov,
  EPJ Web Conf.\  {\bf 192}, 00042 (2018)
  [arXiv:1809.02973 [hep-ph]].

\bibitem{Tanabashi:2018oca} 
  M.~Tanabashi {\it et al.} [Particle Data Group],
  Phys.\ Rev.\ D {\bf 98}, no. 3, 030001 (2018).
  doi:10.1103/PhysRevD.98.030001



  
\end{thebibliography}
